\documentclass[nofootinbib,pra,twocolumn,amsmath,amssymb,aps]{revtex4-1}

\usepackage{graphicx}% Include figure files
\usepackage{dcolumn}% Align table columns on decimal point
\usepackage{bm}% bold math
\usepackage{hyperref}% add hypertext capabilities
\usepackage{color}
\usepackage{physics}

\begin{document}

\title{Manipulating multimer propagation using lattice modulation}
\author{Arya Dhar$^{1}$}
\author{Jami J. Kinnunen$^{2}$}

\affiliation{$^{1}$Institut f\"ur Theoretische Physik, Leibniz Universit\"at Hannover, Appelstra\ss e 2, 30167 Hannover, Germany}
\affiliation{$^{2}$Department of Applied Physics, Aalto University, FI-00076 Aalto, Finland}

\date{\today}
\begin{abstract}
We propose a scheme for controlling the movement of dimers, trimers, and other multimers in optical lattices by modulating the lattice potential. In deep optical lattices the propagation of deeply bound atomic clusters is slowed down by the high energy cost of virtual intermediate states.
Adapting the well-known method of lattice modulation spectroscopy, the movement of the clusters can be made resonant by utilizing sequences of bound-bound transitions. Using the scheme, the mobility of each specific cluster can be selectively controlled by tuning the modulation frequency.
We formulate a simple and intuitive model and confirm the validity of the model by numerical simulations of dimers and trimers in a one-dimensional optical lattice.
\end{abstract}

\maketitle

\section{Introduction}
\label{sec:intro}

Deep optical lattices allow exceptional control of the movement and interactions of individual atoms. The lattice depth determines the wavefunction overlap, directly influencing the on-site and nearest-neighbor interaction strengths and the hopping amplitude between individual lattice sites.
Furthermore, deep optical lattices provide exceptionally long lifetimes by limiting two- and three-body effects, opening interesting opportunities for studying many-body physics but also applications in ultracold chemistry~\cite{ultracoldchemistry_ye,ultracoldchemistry_ospelkaus2010,ultracoldchemistry_ye2017}.

In ultracold chemistry, the formation of molecules is controlled at quantum level. This requires careful control of both spatial and internal degrees of freedom of atoms and molecules~\cite{moleculecreation_ye,moleculecreation_hutson,moleculecreation_cote,moleculecreation_nagerl2014,moleculecreation_cornish2014,moleculecreation_zwierlein2015,moleculecreation_wang2016,moleculecreation_jamison2017}.
One of the outstanding problems is obtaining sufficiently high overlap between the ensemble of individual atoms and the corresponding molecular state. 
This overlap cannot be increased by using higher density of the atomic gases without severely limiting the lifetime of the system. Deep optical lattices provide a way to do both as deeper lattices increase the overlap of atoms in the same lattice site, while at the same time it has the potential of increasing the system lifetime~\cite{latticelifetime_jin2007,latticelifetime_ye2012}.

\begin{figure}[ht]
\centering
\includegraphics[width=0.95\columnwidth]{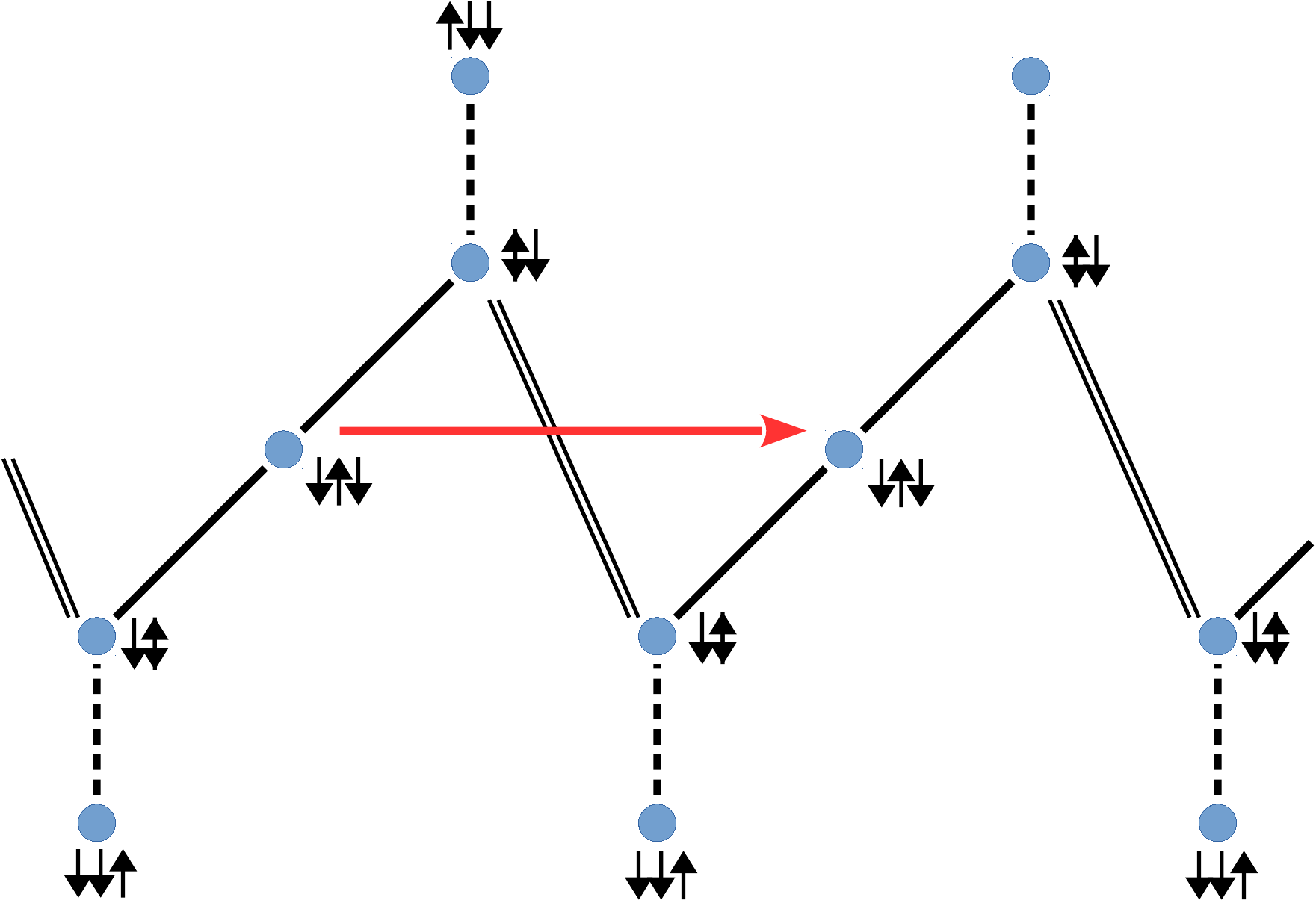}
\caption{Scheme of the lattice modulation driven multimer propagation. Here a trimer propagates resonantly through various bound trimer configurations. 
Red arrow shows the total process of the trimer moving to the adjacent unit cell, corresponding to the hopping of the trimer to the right by one lattice site. This process involves three different internal states $\ket{\downarrow \uparrow \downarrow}$, $\ket{\updownarrow \downarrow}$, and $\ket{\downarrow \updownarrow}$ (first one corresponding to state in which one $\uparrow$-atom is located in a lattice site with two $\downarrow$-atoms in the adjacent sites left and rigth. The latter two refer to states in which the $\uparrow$-atom is in the same site with one $\downarrow$-atom, and the other $\downarrow$-atom is in the adjacent site, right in the first one, left in the second one.) The final state has the same trimer configuration $\ket{\downarrow \uparrow \downarrow}$ as the initial state, but now the trimer has moved to the right by one lattice site. 
The different transitions take place either due to the lattice modulation (solid and dashed lines) or through the usual (non-modulated) hopping (double line). The dashed line transition to the so-called 'dead-end' states is also provided by the lattice modulation, but it will be blocked if there is a mismatch between the intra- and interspecies nearest-neighbor interactions, $V_{\downarrow \downarrow}$ and $V_{\downarrow \uparrow}$.}
\label{fig:trimer_hopping_model}
\end{figure}

At the most microscopic level, the state-of-the-art in ultracold chemistry is controlling the formation of a single individual molecule from atoms trapped by optical tweezers~\cite{ultracoldchemistry_ni}. 
However, deep optical lattices have potential for extending this to ensembles of atoms, assuming that sufficient level of control over the spatial degrees of freedom is obtained~\cite{moleculecreationinlattice_nagerl,moleculecreationinlattice_jin2015,moleculecreationinlattice_ye2019}.

Here we propose a novel scheme for controlling the movement of dimers, trimers and other multimers in deep optical lattice, depicted in Fig.~\ref{fig:trimer_hopping_model}.
Lattice modulation spectroscopy is a well known method~\cite{latticemodulation_esslinger, latticemodulation_inguscio, latticemodulation_tarruell} for studying the spectrum of bound states in optical lattice.
In the method, the amplitude of the lasers that form the lattice is modulated. The generated low-frequency side-bands of these optical fields can form or break pairs of atoms, and the number of doubly occupied sites has generally been the probed signal.
Periodical shaking or modulation of the lattice has also been proposed and used for engineering interesting lattice models~\cite{engineering_dalibard2011,engineering_chin2013,engineering_spielman2014,engineering_santos2014,engineering_santos2016,shakenlattice_esslinger2018}, a concept generally known as Floquet engineering (for review, see~\cite{floquet_review_eckardt2017, Takashi2019}).
In our scheme, the method is adapted for controlling the movement of multimers by driving transitions between specific bound states. 
By choosing proper protocol of bound-bound transitions, the multimer can propagate in the lattice resonantly. 
The method can be used for example for filtering desired multimer configurations or even creating more complex multimers by merging smaller ones.

We formulate an effective multimer propagation model that allows us to estimate the propagation speeds of multimers when driven by resonant lattice modulation. The model is compared with essentially exact numerical simulation of the one-dimensional extended Hubbard model incorporating nearest-neighbor interactions between the atoms. While the focus of the present work is on the control of the propagation of dimers and trimers, the method also provides new possibilities of realizing interesting few- and many-body systems through interactions between the dimers and trimers. 
By controlling the lattice modulation frequency and amplitude, the properties of the multimer excitation spectra can be tuned.
For example, lattice modulation driven dimers and trimers are shown to exhibit a flat band in the excitation spectrum.

The structure of the paper is the following: in Section~\ref{sec:model} we describe the underlying model of extended Hubbard model in one-dimensional lattice. The lattice modulation driven propagation scheme is described in Section~\ref{sec:modulation}, and in Section~\ref{sec:multimer} we provide a simple model of the multimer propagation. We use numerical matrix product state simulation to solve the dynamics according to the extended Hubbard model. Section~\ref{sec:simulation} outlines our numerical simulation and in Section~\ref{sec:comparison} we show comparisons between the numerical simulation and the multimer propagation model. Section~\ref{sec:experiment} discusses briefly the feasibility of realizing this scheme in experiments. Finally, we summarize our key results in Section~\ref{sec:conclusion}.

\section{Model}
\label{sec:model}

Our theory is based on the one-dimensional extended Hubbard model with nearest-neighbor interactions. This model is well-suited for describing low-temperature atoms in optical lattices. The (semi-)long-range interactions can be realized through dipole-dipole interactions or with polar molecules, and all interaction and hopping parameters can be tuned by utilizing interatomic Feshbach resonances and varying the intensity of the lasers creating the optical lattice.

The many-body Hamiltonian describing the system has the form
\begin{eqnarray}
 H &= \sum_{\langle ij\rangle,\sigma} t_0 \hat c^\dagger_{i\sigma} \hat c_{j\sigma} + U\sum_{i} \hat c^\dagger_{i\uparrow} \hat c^\dagger_{i\downarrow} \hat c_{i\downarrow} \hat c_{i\uparrow} \\
 \nonumber &+ \sum_{\langle ij\rangle,\sigma\sigma'} V_{\sigma \sigma'} \hat c^\dagger_{i\sigma} \hat c^\dagger_{j\sigma'} \hat c_{j\sigma'} \hat c_{i\sigma}
\end{eqnarray}
where $U$ is the on-site interaction strength, $V_{\sigma \sigma'}$ is the nearest-neighbor interaction strength for two atoms in spin states $\sigma$ and $\sigma'$, and the operator $\hat c_{i\sigma}^{(\dagger)}$ removes (creates) a fermionic atom of spin $\sigma$ in lattice site $i$. The movement of the atoms in the lattice is described by the hopping term with magnitude $t_0$, which is here assumed to be the same for both spin states. The summation over $\langle ij\rangle$ refers to summation over adjacent sites for which $j=i+1$ or $j=i-1$.
Here we assume only two spin states $\downarrow$ and $\uparrow$, but the model can be extended to more complicated systems as well, and the different spin states may also refer to atoms of different species.
The model neglects direct multiparticle hopping and the movement of multimers thus takes place through sequences of single-particle hoppings. 
%{\color{red} Can we find some references for this? From Luis Santos?}

In all of the calculations in this paper, we consider only few-particle systems with either one $\uparrow$-atom and one $\downarrow$-atom (forming a dimer), or one $\uparrow$-atom and two $\downarrow$-atoms (forming a trimer). Notice that for these calculations, possible $\uparrow$-$\uparrow$ interaction is irrelevant.
The interaction parameters that we use throughout the paper are $U=-40\,t_0$, $V_{\downarrow \uparrow} = V_{\downarrow \downarrow} = -15\,t_0$.
Ignoring the small effect of hopping, for these parameters the trimer state has bound state energies of $-70\,t_0$ (for trimer configurations $\ket{\updownarrow \downarrow}$ and $\ket{\downarrow \updownarrow}$, that is, two $\downarrow$-atoms are in adjacent sites and the $\uparrow$ atom is in the same site with either forming a doublon $\updownarrow$) and $-30\,t_0$ (for configuration $\ket{\downarrow \uparrow \downarrow}$, $\ket{\uparrow \downarrow\downarrow}$, and $\ket{\downarrow \downarrow \uparrow}$). Bound dimer states have energies $-40\,t_0$ (for on-site dimer $\ket{\updownarrow}$) and $-15\,t_0$ (for nearest-neighbor dimers $\ket{\downarrow \uparrow}$ and $\ket{\uparrow \downarrow}$). This chosen set of parameters is not special, but it allows us to utilize various bound-bound transitions without interfering with other transitions. For example, bound trimer transition can be driven resonantly with lattice modulation of frequency $\omega = 40\,t_0$ and the bound dimer transition with frequency $\omega=25\,t_0$. Neither of these frequencies is resonant with any of the multimer-breaking processes (for example, trimers in different configurations can be broken at frequencies $10\,t_0$, $15\,t_0$, $30\,t_0$, $55\,t_0$, and $70\,t_0$). Experimentally these bound state energies can be tuned by varying interaction strengths using for example Feshbach resonances and changing the lattice depth.

The lattice modulation is described by the operator
\begin{equation}
    V_\mathrm{m} = \sum_{\langle ij\rangle,\sigma} t_\mathrm{m} \sin \left(\omega t\right) \hat c_{i\sigma}^\dagger \hat c_{j\sigma}
\end{equation}
where $t_\mathrm{m}$ is the lattice modulation amplitude and $\omega$ the frequency of the modulation.
This operator can be understood to provide an additional way for the atoms to hop between adjacent sites by absorbing or emitting photons of frequency $\omega$ into the modulating field. 
By properly choosing the modulation frequency $\omega$, single-particle hopping transitions that would otherwise be blocked due to different binding energies can be allowed.

The lattice modulation operator is same as used in lattice modulation spectroscopy. However, the difference here is that the modulation is not used primarily for probing properties of the underlying system but for providing it new properties. Indeed, the modulation is assumed to be present for arbitrarily long time, during which the system may evolve far from the initial state. 

The form of the lattice modulation operator is strictly valid only in the limit of weak modulation amplitude $t_\mathrm{m}$.
This is because the amplitude modulation is realized through modulation of the lattice potential, but the hopping parameter $t$ and the interaction parameters $U$ and $V$ have non-linear dependence on the lattice depth $V(x)$~\cite{latticeparameters_jaksch1998,latticeparameters_zwerger2003}.
Still, most of the calculations below are done assuming the simple sinusoidal form of the lattice modulation even for relatively large modulation amplitudes.
This is required in order to reduce the numerical complexity as lower modulation amplitudes are reflected as slower propagation speeds, which in turn would require longer simulation times.
In an experiment, the modulation amplitude can be lower as slower propagation speeds are not a problem assuming lifetimes are sufficiently long.

\section{Lattice modulation driven propagation}
\label{sec:modulation}

Attractive nearest-neighbor interactions provide several bound states for dimers and trimers. The goal of the lattice modulation driven propagation is to utilize these bound states and realize the multimer propagation through bound-bound transitions.
A sample process is shown for the trimer propagation in Fig.~\ref{fig:trimer_hopping_model}. In the strongly interacting limit $|U|,|V| \gg t_0$ there are five bound trimer configurations (only three in the absence of intraspecies nearest-neighbour interactions): $\ket{\downarrow \downarrow \uparrow}$, $\ket{\downarrow \updownarrow}$, $\ket{\downarrow \uparrow \downarrow}$, $\ket{\updownarrow \downarrow}$, and $\ket{\uparrow \downarrow \downarrow}$. The movement of the trimer to the right by a single lattice site (shown as the red arrow) takes place by a sequence of bound-bound transitions. Two of the transitions $\ket{\downarrow \updownarrow} \rightarrow \ket{\downarrow \uparrow \downarrow}$ and $\ket{\downarrow \uparrow \downarrow} \rightarrow \ket{\updownarrow \downarrow}$ involve the lattice modulation term of the Hamiltonian, whereas the last transition $\ket{\updownarrow \downarrow} \rightarrow \ket{\downarrow \updownarrow}$ involves simple hopping of the $\uparrow$-atom. 

As seen in the model, the frequency of the lattice modulation should match the energy separation between the trimer configurations $\ket{\downarrow \uparrow \downarrow}$ and $\ket{\updownarrow \downarrow}$. These have the binding energies of $2V_{\downarrow \uparrow}$ and $U+2V_{\downarrow \uparrow}$, respectively, yielding the energy separation equal to $U$. Notice that this frequency is far from any trimer-breaking transitions and hence the trimer propagation can be realized using only bound trimer configurations.

A few illuminating observations can be already seen from the model of Fig.~\ref{fig:trimer_hopping_model}. First of all, in the case that intra- and interspecies nearest-neighbor interactions have sufficient mismatch and the two 'dead-end' states $\ket{\uparrow \downarrow \downarrow}$ and $\ket{\downarrow \downarrow \uparrow}$ can be neglected, the trimer propagation model yields a simple linear chain. In the case of lattice modulation amplitude being equal to the hopping amplitude $t_\mathrm{m} = t_0$, the propagation of the trimer by a single lattice site would correspond to a single particle propagating in a linear chain by three sites. That is, the trimer propagation speed becomes one third of the single particle propagation speed. 
In practice, though, this cannot be achieved as the sinusoidal modulation amplitude $t_\mathrm{m}$ is always less than the bare hopping amplitude $t_0$.

Another interesting observation is that, in the case of equal intra- and interspecies nearest-neighbor interactions $V_{\downarrow \downarrow} = V_{\downarrow \uparrow}$, the trimer spectrum should have a flat band.
This comes from the possibility of having a superposition of states $\ket{\downarrow \uparrow \downarrow}$, $\ket{\uparrow \downarrow \downarrow}$, and $\ket{\downarrow \downarrow \uparrow}$. With equal probabilities of all these configurations and opposite sign for the $\ket{\downarrow \uparrow \downarrow}$ configuration, the hopping to the states $\ket{\updownarrow \downarrow}$ and $\ket{\downarrow \updownarrow}$ are blocked by complete destructive interference. This effect does not depend on the magnitude of the hopping modulation amplitude $t_\mathrm{m}$ nor even the bare hopping amplitude $t_0$.

\begin{figure}[ht]
\centering
\includegraphics[width=0.95\columnwidth]{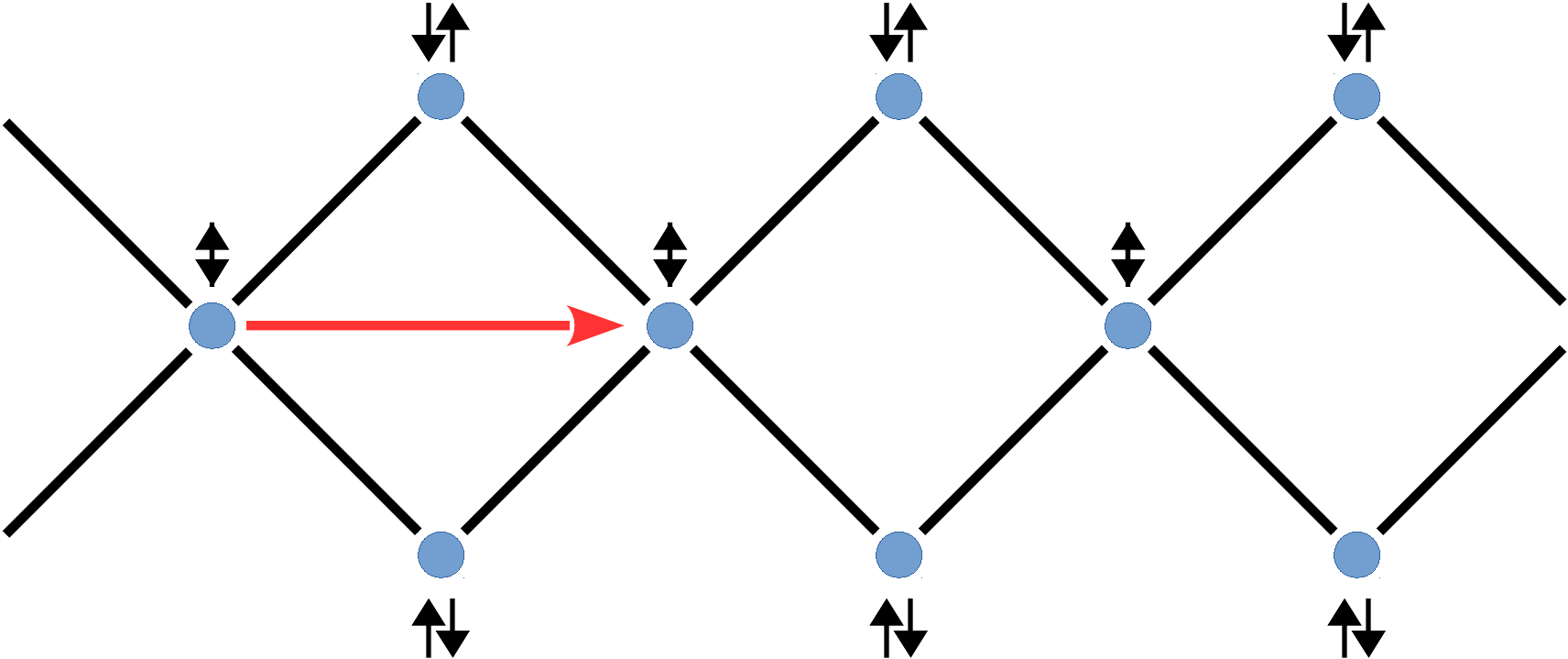}
\caption{Model of the nearest-neighbor dimer propagation through various bound dimer configurations. Red arrow shows the total process of the dimer hopping to the right by one site. This process involves two alternative paths involving internal states $\ket{\downarrow \uparrow}$ or $\ket{\uparrow \downarrow}$.}
\label{fig:dimer_hopping_model}
\end{figure}

The flat-band dispersion is even easier to notice in the case of nearest-neighbor dimer propagation model. Fig.~\ref{fig:dimer_hopping_model} shows the diamond chain lattice of the dimer propagation. This (effective) lattice is known to have a flat band~\cite{flatband_vidal2000,flatbands_huber}. The unit cell has three sites and hence the nearest-neighbor dimer has three excitation bands and one of them is flat while the other two are dispersive.

\section{Effective multimer propagation model}
\label{sec:multimer}

The simple model of the lattice modulation driven propagation allows us to write an effective multimer hopping Hamiltonian and solve the dispersion relations of dimers and trimers (and in principle also larger multimers).
The effective multimer hopping Hamiltonian has the generic form
\begin{equation}
    H = \sum_{i,\sigma,\sigma'} K^0_{\sigma,\sigma'} \hat m_{i\sigma}^\dagger \hat m_{i\sigma'} + \sum_{i,\sigma,\sigma'} K^1_{\sigma,\sigma'} \hat m_{i+1,\sigma}^\dagger \hat m_{i\sigma'} + H.c.,
    \label{eq:multimer_hopping_H}
\end{equation}
in which the operator $\hat m_{i\sigma}^{(\dagger)}$ annihilates (creates) the multimer in site $i$ and bound-state configuration $\sigma$. The summation over $\sigma$ is over all multimer configurations that have the same binding energy or are connected resonantly through the lattice modulation. 
The first term in Eq.~\eqref{eq:multimer_hopping_H} describes transitions between various internal states of the multimer within the unit cell caused by hoppings of the individual atoms. Also the second term comes from a hopping of an individual atom, but here the transition is to a multimer state that belongs to the adjacent unit cell, thus describing the hopping of the multimer to the adjacent lattice site.

Diagonal elements of the matrix $K^0$ correspond to the binding energies of the multimer configurations. However, in the rotating-wave approximation of the lattice modulation, the modulation frequency $\omega$ can be incorporated as a constant energy shift of the coupled energy levels.
Assuming that the modulation frequency is chosen to match the energy separation of the bound states, the diagonal elements of the matrix $K^0$ become equal and can be removed by simple energy shift.
This makes the lattice modulation operator time-independent, leading into time-independent coupling terms in matrices $K^0$ and $K^1$~\cite{rotatingwaveapproximation_torma}.

\subsection{Nearest-neighbor dimer model}

As an example, for the nearest-neighbor dimer model shown in Fig.~\ref{fig:dimer_hopping_model}, the relevant internal states are the two nearest-neighbor dimers $\ket{\uparrow \downarrow}$ and $\ket{\downarrow \uparrow}$, and one state for the on-site dimer $\ket{\updownarrow}$. The corresponding on-site matrix is
\begin{equation}
   K^0 = \left(\begin{array}{ccc}
        0 & t_\mathrm{m} & 0 \\
        t_\mathrm{m} & 0 & t_\mathrm{m} \\
        0 & t_\mathrm{m} & 0
   \end{array}\right),
\end{equation}
where the middle row corresponds to the on-site dimer and the upper and lower rows to the nearest-neighbor dimer configurations.
While the on-site term of the model Hamiltonian does not provide actual hopping of the multimer but merely rotations in the internal state space, the nearest-neighbor hopping term allows the actual propagation. For the nearest-neighbor dimer this is given by the matrix
\begin{equation}
   K^1 = \left(\begin{array}{ccc}
        0 & 0  & 0 \\
        t_\mathrm{m} & 0 & t_\mathrm{m} \\
        0 & 0 & 0
   \end{array}\right),
\end{equation}
which describes the process in which the dimer in either of the the nearest-neighbour dimer configuration is coupled to the on-site dimer configuration on the adjacent site.
For trimers, the number of internal states is larger, and the corresponding hopping matrices more complicated, but still the model Hamiltonian satisfies the generic form of Eq.~\eqref{eq:multimer_hopping_H}, see Appendix A.

\subsection{Multimer dispersion}

\begin{figure}[ht]
    \centering
    \includegraphics[width=0.95\columnwidth]{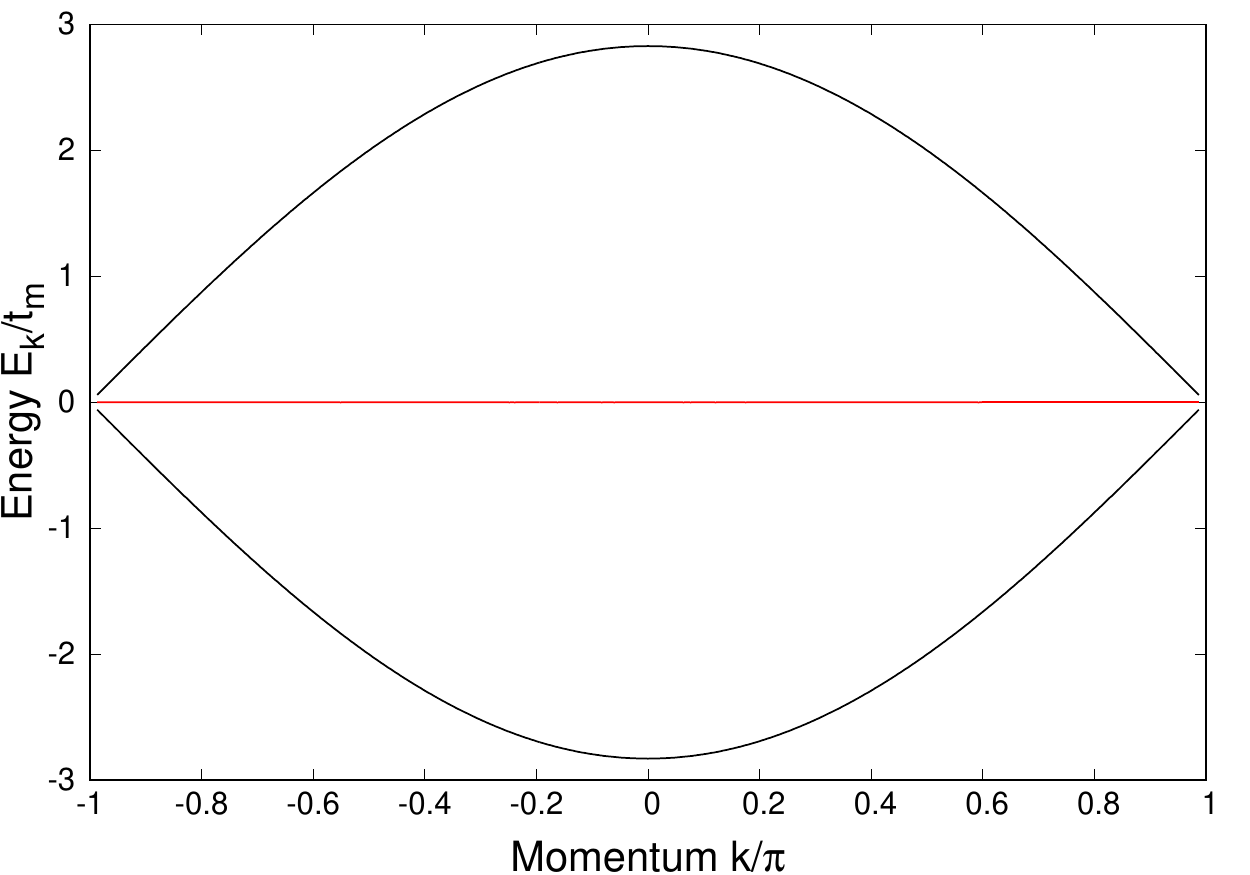}
    \caption{Nearest-neighbor dimer dispersion with resonant lattice modulation. Excitation spectrum shows three bands, with two dispersive bands (shown in black) and one flat band (in red). Notice that y-axis is scaled by the modulation amplitude $t_\mathrm{m}$ instead of bare hopping $t_0$. The bare hopping $t_0$ does not enter the dimer model and hence the amplitude $t_\mathrm{m}$ is the only energy scale in the model.}
    \label{fig:dimer_dispersion}
\end{figure}

The dispersion of the multimer configuration in this resonant model can be calculated by taking the Fourier transform of the effective Hamiltonian yielding
\begin{equation}
    H = \sum_{k,\sigma,\sigma'} \left[K^0_{\sigma,\sigma'} + e^{-i k} K^1_{\sigma,\sigma'} + e^{i k} K^{1\dagger}_{\sigma,\sigma'}\right] \hat m_{k\sigma}^\dagger \hat m_{k\sigma'},
\end{equation}
where $k$ are dimensionless momenta $k \in \left[-\pi,\pi\right]$.
The dispersion relations are thus obtained from the eigenvalues of the matrix
\begin{equation}
M = K^0 + e^{-ik}K^1 + e^{i k}K^{1\dagger}
\end{equation}
The dimensionality of the matrix $M$ determines the number of bands. In the case of nearest-neighbor dimer model described above, the matrix is
\begin{equation}
M = \left(\begin{array}{ccc}
        0 & t_\mathrm{m}(1+e^{-ik}) & 0 \\
        t_\mathrm{m}(1+e^{i k}) & 0 & t_\mathrm{m}(1+e^{ik}) \\
        0 & t_\mathrm{m}(1+e^{-ik}) & 0
   \end{array}\right).
\end{equation}
Eigenvalues of this matrix are $\lambda = 0$ and $\lambda = \pm 2\sqrt{2}t_\mathrm{m} \cos \left(k/2\right)$. Dimer dispersion is plotted in Fig.~\ref{fig:dimer_dispersion} showing one flat and two dispersive bands.

\begin{figure}[ht]
    \centering
    \includegraphics[width=0.95\columnwidth]{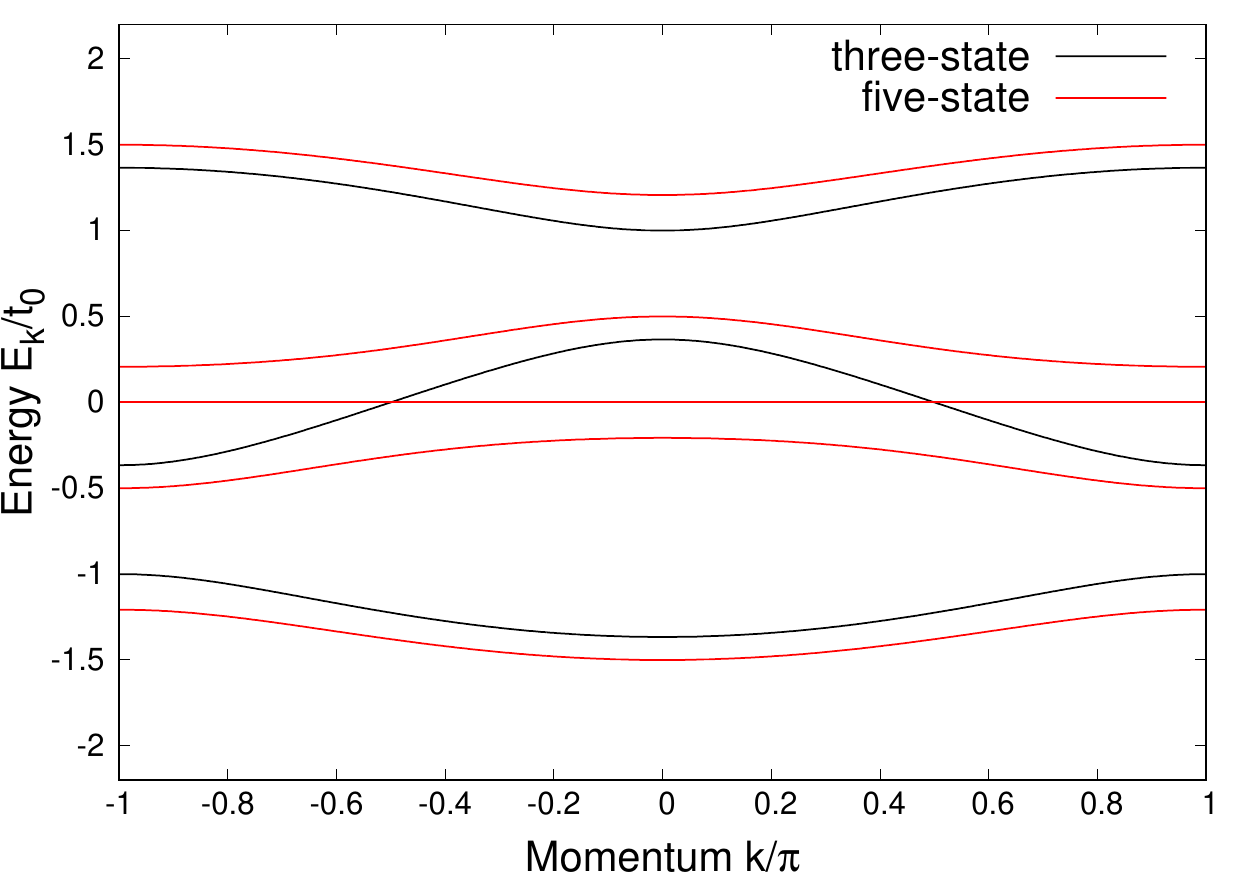}
    \caption{Trimer excitation spectrum for hopping modulation amplitude $t_\mathrm{m}=0.5\,t_0$. Shown are the three bands of the three-state model (in black) and five bands of the five-state model (in red) that includes the coupling to dead-end states. The third (middle) band in the five-state model is flat.}
    \label{fig:trimer_dispersion}
\end{figure}

Fig.~\ref{fig:trimer_dispersion} shows the excitation spectrum of a trimer for two cases: three-state trimer corresponds to the model in which the intra- and interspecies nearest-neighbor interactions have large offset, effectively blocking the dead-end states in Fig.~\ref{fig:trimer_hopping_model}. In contrast, the five-state trimer corresponds to symmetric interaction $V_{\downarrow \downarrow} = V_{\downarrow \uparrow}$, making also the lattice modulation transitions to dead-end states resonant.
The figure shows the band structures of the two cases, and the flat band in the five-state model is clearly seen as the middle band at zero energy.

\subsection{Multimer propagation speed}

The speed of the multimer propagation is determined by the dispersion relation $\epsilon_k$ through the relation
\begin{equation}
    v = \frac{\partial \epsilon_k}{\partial k}.
    \label{eq:speed_and_dispersion}
\end{equation}
As different excitation bands have clearly different dispersion relations, the propagation speed will depend both on the band index $\sigma$ and the momentum $k$. If the initial state of the multimer wavefunction is localized, the momentum $k$ is not well defined and the observed propagation speed is expected to be an average of Eq.~\eqref{eq:speed_and_dispersion} calculated over the Brillouin zone.
This yields as the average propagation speed for multimer in band $\sigma$
\begin{equation}
    v_\mathrm{avg}^\sigma = \frac{1}{\pi}\left|\epsilon_{\pi}^\sigma - \epsilon_0^\sigma \right| = \frac{\Delta_\sigma}{\pi},
    \label{eq:average_speed}
\end{equation}
where $\Delta_\sigma$ is the width of the excitation band $\sigma$.

For the nearest-neighbor dimer, the dispersion relation was solved above and the width of the two dispersive bands is $2\sqrt{2} t_\mathrm{m}$, which scales linearly with modulation amplitude $t_\mathrm{m}$. The linear scaling might be surprising, considering that the hopping of the dimer by one site involves two processes with amplitude $t_\mathrm{m}$ each. However, the process does not involve virtual intermediate states as the lattice modulation makes the transitions resonant. 
Indeed, one can view the lattice modulation as causing Rabi oscillations between the bound dimer states.
However, now the Rabi oscillations are associated with the hopping process, and hence the propagation speed scales in the same way as resonant Rabi frequency, which has linear dependence on the coupling constant.

%Instead, as the dimer model in Fig.~\ref{fig:dimer_hopping_model} shows, the hopping of the dimer to the adjacent site involves two steps, making the lattice effectively twice as long. Secondly, there are two possible paths, which further increases the probability of hopping, resulting in propagation speed comparable to the speed of a single particle in the lattice. 

\begin{figure}[ht]
    \centering
    \includegraphics[width=0.49\columnwidth]{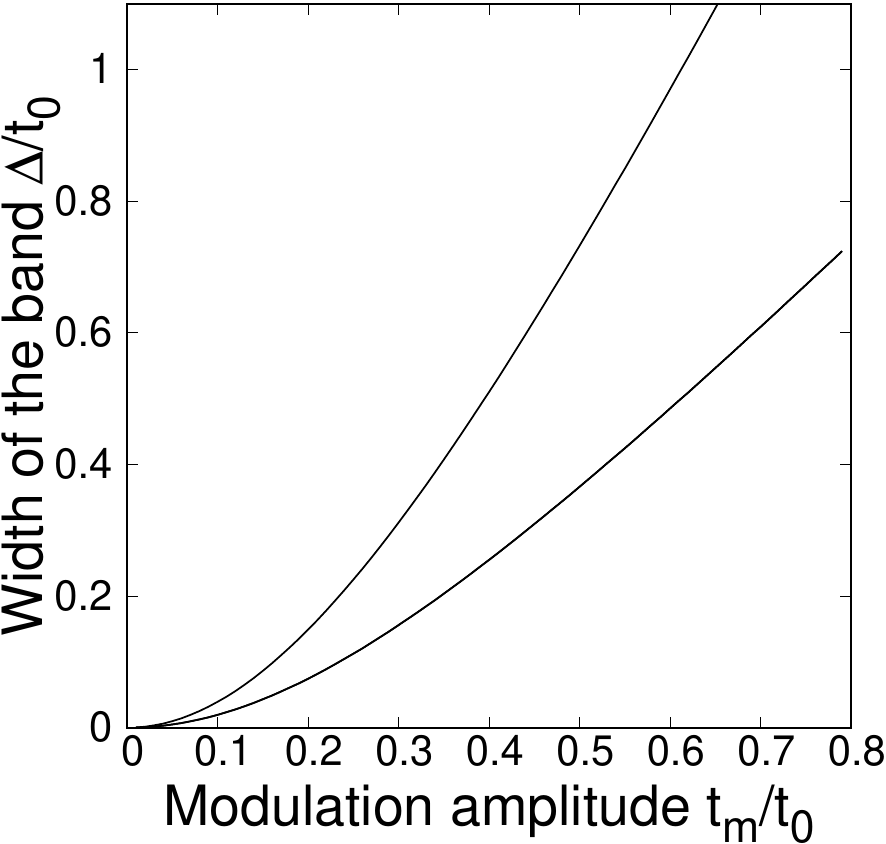}
    \includegraphics[width=0.49\columnwidth]{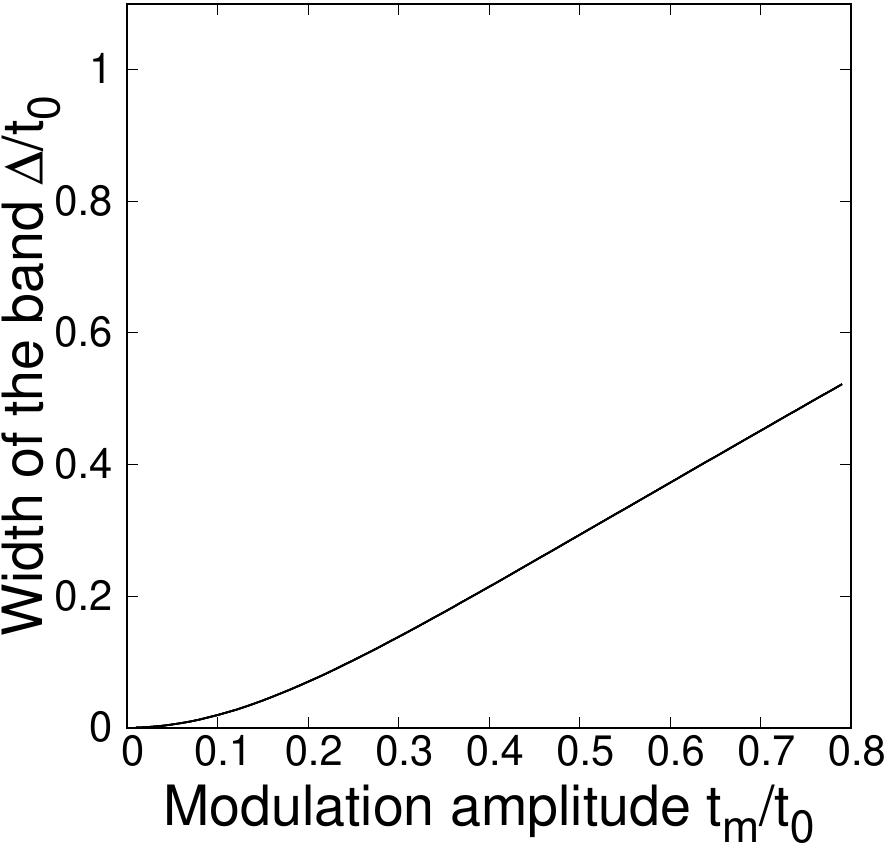}  
    \caption{Widths of the trimer excitation bands. For three-state trimer (left), two of the bands have equal widths (lower curve) and the third band has width equal to twice the width of the other two. 
    For the five-state trimer (right), four of the bands have equal widths, while one band is flat with zero width.}
    \label{fig:trimer_dispersion_width}
\end{figure}

Figure~\ref{fig:trimer_dispersion_width} shows how widths of the various trimer bands increase rapidly with the modulation amplitude.
The three-state model incorporates two narrow bands (top and bottom bands in Fig.~\ref{fig:trimer_dispersion}) and one band with double the width. 
These widths can be calculated analytically.
The two narrower bands have widths
\begin{equation}
    \Delta = 2t_0\sqrt{\frac{2r^2+1}{3}} \left[ \cos \frac{\phi}{3} - \cos \left(\frac{\pi}{3}-\frac{\phi}{3}\right)\right],
    \label{eq:threestatewidth}
\end{equation}
where $\phi = \arctan \sqrt{\left(\frac{2r^2 + 1}{3}\right)^3 \frac{1}{r^4} - 1}$, and $r = t_\mathrm{m}/t_0$. The third band has width exactly twice the width of the other two.
In the limit of weak lattice modulation, $r \rightarrow 0$, Eq.~\eqref{eq:threestatewidth} yields $2t_\mathrm{m}^2/t_0$.

The five-state trimer model, which includes dead-end states,
has four bands with equal widths and one flat band of zero width. The non-flat bands yield slightly lower propagation speeds than the three-state model. 
This is a combinatorial effect since additional states will reduce the probability that the trimer will eventually complete the whole cycle of transitions required for the hopping process. 
For the five-state trimer the reduction in speed is roughly 50\% compared to the three-state trimer. However, the combinatorial slow-down may turn out to be much more important in the case of larger multimers, if the number of dead-end states increases significantly with the larger number of bound states. On the other hand, as seen in the case of three-state model, it may be possible to eliminate these states by breaking symmetries in interaction channels.

\section{Numerical simulations}
\label{sec:simulation}

We have performed numerical simulations of the lattice modulation driven propagation of dimers and trimers in one-dimensional lattice using the highly accurate method of the matrix product states (MPS)~\cite{Wall_2012}. MPS has proved to be a very successful numerical technique in describing both static and dynamic properties of one dimensional systems~\cite{Schollwock2011}. 

The system is initialized by solving the interacting ground state in a small box of two sites at the center of a larger lattice of 100 sites for both dimers and trimers. 
At time $t=0$ the lattice modulation is switched on, the box is opened and the atoms expand into the larger lattice. The time evolution is evaluated using the time-dependent variational principle (TDVP)~\cite{Verstraete_2016} to calculate the time evolution of the density distribution of both $\uparrow$ and $\downarrow$ atoms. The time step in the dynamics was fixed to $0.01\,\frac{1}{t_0}$. However we have performed convergence checks for lower time step values. The ground state calculations have been performed with a bond dimension of 1000, whereas for the time evolution, up to 500 states have been kept leading to an error of less than $10^{-8}$.

The numerical results below show the $\downarrow$-atom density. Since the lattice modulation frequency is far from any dimer- or trimer-breaking transitions, considering only the single-particle density is sufficient for determining the propagation of the multimer.

\section{Comparison with simulations}
\label{sec:comparison}

\begin{figure}[ht]
    \centering
    \includegraphics[width=0.95\columnwidth]{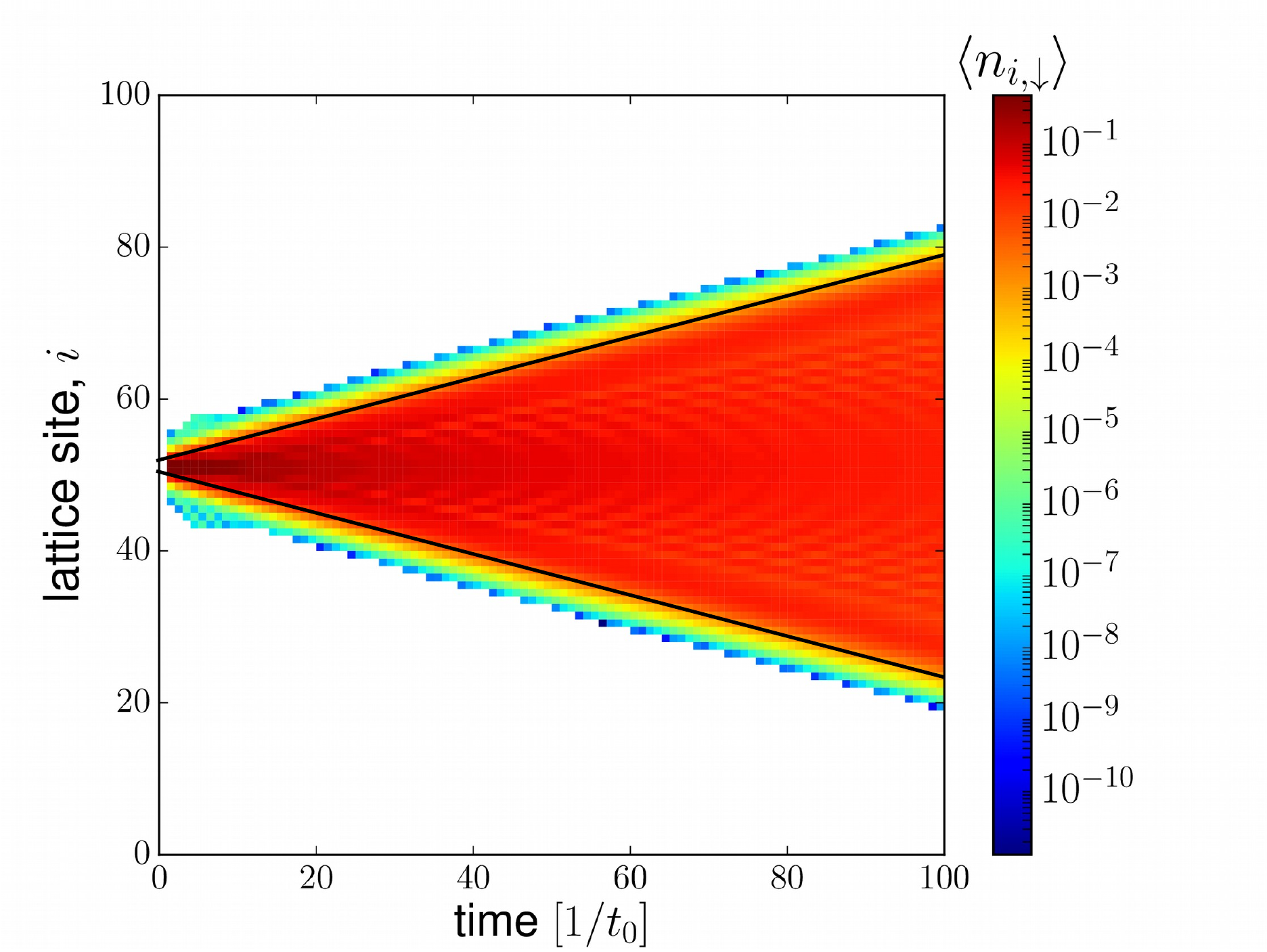}
    \includegraphics[width=0.95\columnwidth]{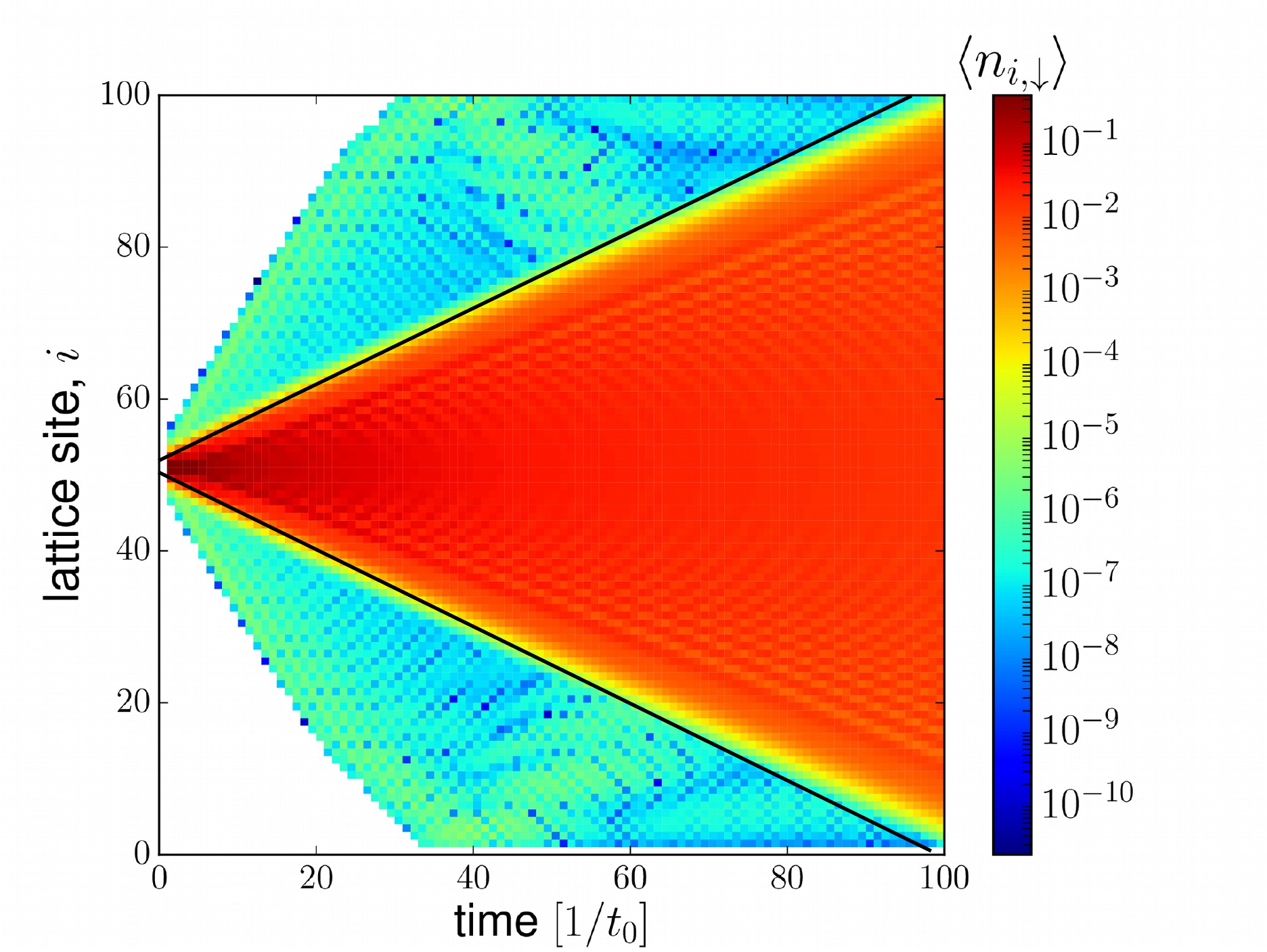}
    \caption{Numerical simulation of the nearest-neighbor dimer propagation with lattice modulation (modulation amplitude $t_m = 0.3\,t_0$ in the top and $t_m = 0.6\,t_0$ in the lower figure). 
    Solid black lines show the expected propagation speeds $\frac{2\sqrt{2}t_\mathrm{m}}{\pi}$. Numerical data shows the logarithmic $\downarrow$-atom density $\langle n_{i,\downarrow}\rangle$ as a function of time and position.}
    \label{fig:dimer_propagation_numerics}
\end{figure}

Figure~\ref{fig:dimer_propagation_numerics} shows the simulated propagation of the nearest-neighbor dimer for two different modulation amplitudes. The frequency of the modulation, $\omega = V_{\downarrow \uparrow}-U=25\,t_0$, is at the resonance of the bound-bound transition between deeply bound on-site dimer (with binding energy $U=-40\,t_0$) and nearest-neighbor dimer (with binding energy $V_{\downarrow \uparrow}=-15\,t_0$). The simulation thus satisfies the assumptions of the dimer model in Fig.~\ref{fig:dimer_hopping_model}.
Figure~\ref{fig:dimer_propagation_numerics} shows also the expected density wavefronts based on the propagation speed calculated from the model. The numerical and analytical results agree very well, confirming the predicted linear scaling of the propagation speed.

\begin{figure}[ht]
    \centering
    \includegraphics[width=0.98\columnwidth]{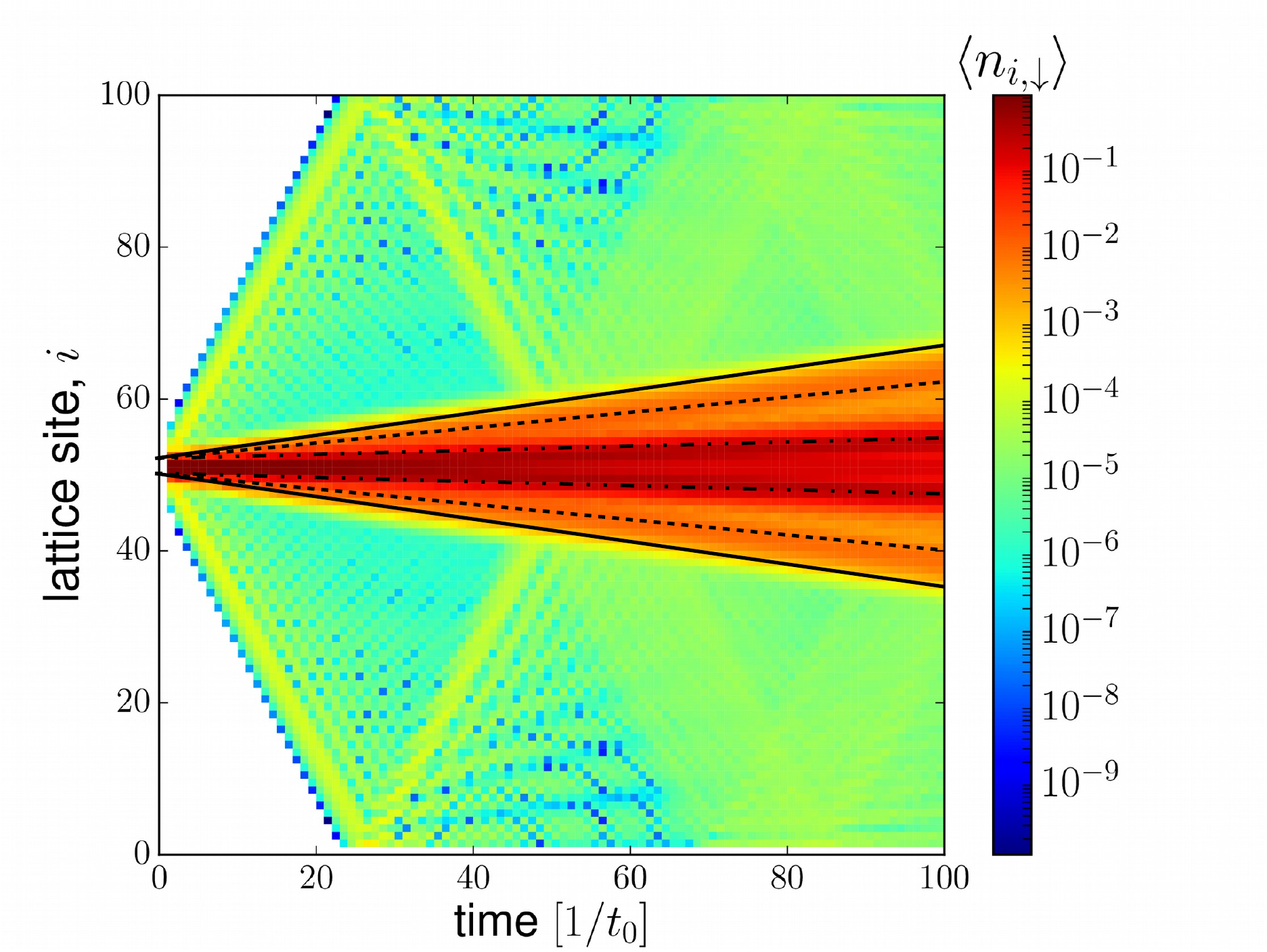}
    \includegraphics[width=0.98\columnwidth]{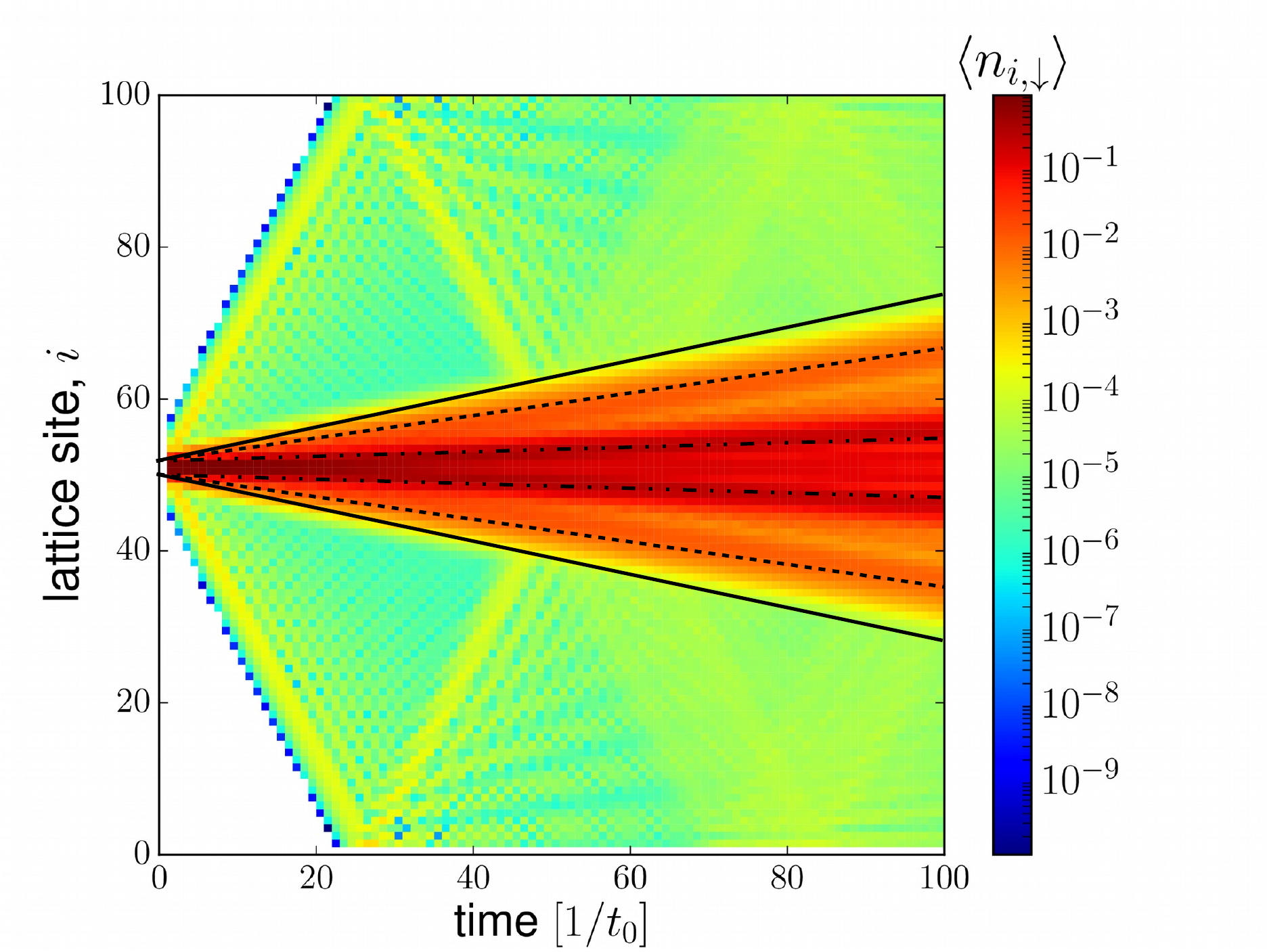}
    \caption{Numerical simulation of the trimer propagation with lattice modulation. Modulation amplitudes are $t_\mathrm{m}=0.5\,t_0$ (top) and $t_\mathrm{m} = 0.7\,t_0$ (bottom). Solid black line shows the maximum velocity in the dispersion spectrum according to Eq.~\eqref{eq:speed_and_dispersion}. Dashed line shows the average speed from Eq.~\eqref{eq:average_speed}, and the dash-dotted line shows the expected propagation in absence of lattice modulation. Numerical data shows the logarithmic $\downarrow$-atom density $\langle n_{i,\downarrow}\rangle$ as a function of time and position.}
    \label{fig:trimer_propagation_numerics}
\end{figure}

\begin{figure}[ht]
    \centering
    \includegraphics[width=0.98\columnwidth]{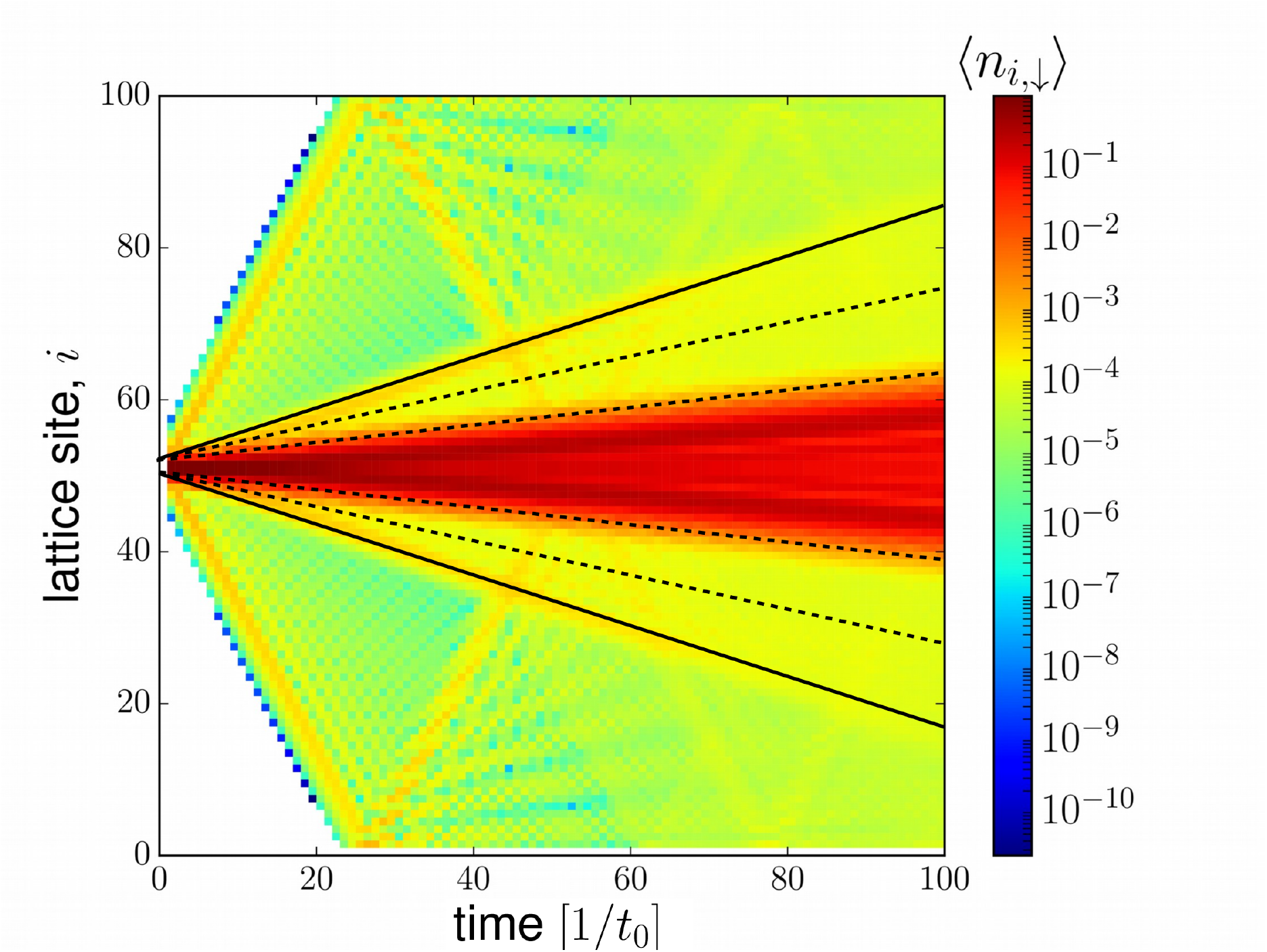}
    \caption{Numerical simulation of the three-state trimer propagation with lattice modulation. Modulation amplitude is $t_\mathrm{m} = 0.5\,t_0$. Solid black line shows the maximum velocity from Eq.~\eqref{eq:speed_and_dispersion} while the two dashed black lines show the average propagation speeds of the two bands. Numerical data shows the logarithmic $\downarrow$-atom density as a function of time and position. Compared to Fig.~\ref{fig:trimer_propagation_numerics}, here the intraspecies nearest-neighbor interaction $V_{\downarrow \downarrow}$ is set to zero, eliminating the dead-end states of the five-state model.}
    \label{fig:trimer_threestate_propagation_numerics}
\end{figure}

Figure~\ref{fig:trimer_propagation_numerics} shows the simulated propagation of a trimer with equal intra- and interspecies nearest-neighbor interactions, $V_{\downarrow \downarrow}=V_{\downarrow\uparrow}$.
The lattice modulation is driven at frequency $\omega =|U|=40\,t_0$. This corresponds to the transition between deeply bound trimer states, and thus satisfies the assumptions of the five-state trimer model in Fig.~\ref{fig:trimer_hopping_model}.
The figure shows the wavefront-like spreading of the initially localized trimer wavepacket. 
Overlaid in the figures are the expected maximum propagation speed determined as the maximum of Eq.~\eqref{eq:speed_and_dispersion} for the calculated dispersion and the average of the propagation speed from Eq.~\eqref{eq:average_speed}. The model fits the results from the simulation very well.
Furthermore, shown is also the expected propagation speed in absence of the lattice modulation. This describes the flat band contribution, as the hopping of the $\uparrow$ and $\downarrow$ atoms involving virtual broken-trimer states yield finite propagation speed even for the flat band states. In the limit of very deep optical lattice, this contribution also vanishes.

Figure~\ref{fig:trimer_threestate_propagation_numerics} shows the similar simulated trimer propagation as in Fig.~\ref{fig:trimer_propagation_numerics}. However, here the intraspecies nearest-neighbor interaction $V_{\downarrow \downarrow}$ is set to zero. This means that two of the trimer configurations $\ket{\downarrow \downarrow \uparrow}$ and $\ket{\uparrow \downarrow \downarrow}$ become far off-resonant from the lattice modulation and the system thus realizes the three-state trimer model of Fig.~\ref{fig:trimer_hopping_model}. Predicted propagation speeds can again be seen to describe well the observed time evolution.

\section{Experimental considerations}
\label{sec:experiment}

The lattice modulation scheme proposed here is  applicable to present experiments done with dipolar and polar gases in optical lattices. 
Deep optical lattice yields long lifetimes and also opens wide band gaps to higher lattice bands suppressing interband transitions. Low temperatures are not required, and the modulation itself is a well-known method. A closely related experiment in 2011 used lattice modulation together with a tilted lattice to control propagation of bosonic atoms in a strongly interacting Mott-insulator state~\cite{floquet_experiment_greiner2011}.

The challenge of the present scheme is finding suitable combinations of on-site and nearest-neighbor interactions.
The method requires having large separations between various bound-bound transitions. 
However, part of the bound states may be even repulsively bound~\cite{repulsivelyboundpairs_zoller}.
For example, attractive nearest-neighbor interaction $V_{\downarrow \uparrow} = V_{\downarrow\downarrow}$ and repulsive on-site  interaction $U$ would give trimer states with energies $2V$ (for configurations $\ket{\downarrow \uparrow\downarrow}$, $\ket{\uparrow \downarrow \downarrow}$ and $\ket{\downarrow \downarrow \uparrow}$) and $2V-U$ (for configurations $\ket{\updownarrow \downarrow}$ and $\ket{\downarrow \updownarrow}$). 
For most values of $U$ and $V$, the bound-bound trimer transition at frequency $\omega = |U|$ is far from any trimer-breaking transitions. 
This is true even in the case of $U>|2V|$, which corresponds to a transition between the ground state trimer and a repulsively bound trimer state with energy larger than zero.

In an experimental realization, the interaction parameters $U$ and $V$ should be even larger relative to the hopping $t_0$ than used in this work. For example, the bare propagation speed of dimers in absence of lattice modulation scales as $4t_0^2/(U-V)$, which for the parameters used here is close to half of the propagation speed seen in Fig.~\ref{fig:dimer_propagation_numerics}. Even deeply bound trimers propagate in absence of lattice modulation, albeit generally at much slower speed~\cite{fasttrimers}. 
In order to have significant speed-up from the lattice modulation, the bare propagation speed needs to be lower.
Fortunately, increasing lattice depth will strongly suppress hopping $t_0$, thus increasing ratios $U/t_0$ and $V/t_0$.
However, care must be taken to avoid various molecular states that may arise from confinement-induced resonances. For that purpose having repulsive on-site interaction $U$ may be advantageous.

If the nearest-neighbor interaction is large, then the next-nearest-neighbor interactions may also be significant. 
While this makes the bound state structure of the system more complicated, additional bound states may improve the scheme, as more options will be available for realizing the hopping protocol.

Experimental setup realizing the scheme could involve an expansion in a lattice and filtering out all singlons and trimers in order to obtain a system consisting of dimers only. By modulating the lattice to enhance the mobility of trimers, with singlons always moving relatively rapidly, only dimers will be left in the initial position. Similarly a purified trimer-gas could be generated by lattice modulation at bound-bound dimer transitions.
Resulting multimer gases can then be detected using quantum gas microscopes~\cite{microscope_thywissen2015, microscope_gross2015, microscope_zwierlein2015, microscope_kuhr2015, microscope_greiner2015, microscope_bakr2017} or with lattice modulation spectroscopy.

\section{Conclusion}
\label{sec:conclusion}

We have studied lattice modulation driven propagation of dimers and trimers in a simple one-dimensional lattice. 
Nearest-neighbor interactions provide multiple bound states for the multimers. 
Resonantly driving bound-bound transitions, the propagation of multimers in a deep optical lattice can be controlled. 
A simple effective model was formulated and shown to provide an accurate description of the scheme.

Resonant lattice modulation was shown to provide also interesting dispersion relations, with flat bands appearing in nearest-neighbor dimer and trimer excitation spectra. The dispersion relations can further be varied by detuning the modulation frequency, using multiple frequencies (multicolor modulation), and by having different modulations for different atomic species.

The propagation of multimers can further be controlled by tilting the lattice. Combining lattice modulation with the tilt, even the propagation direction can be chosen. 

And finally, this work considered only one-dimensional optical lattice. However, the scheme itself is applicable to two- and three-dimensional lattices as well, and also lattices with more complicated geometry, as long as suitable bound-bound transition protocols can be found for enhanced propagation.

%{\color{red} Maybe some words about: nonsinusoidal modulation (true modulation). What if off-resonant?}
%{\color{red}If we modulate $V(x)$ with appreciable amplitude, how should $U$ and $V$'s change?}

\begin{acknowledgments}
We would like to thank Jani-Petri Martikainen and Sebastiano Peotta for good suggestions and interesting discussions. A.D. would like to thank Daniel Jaschke for interesting discussions regarding open source matrix product states (OSMPS) method.
\end{acknowledgments}

\bibliographystyle{apsrev4-1}
\bibliography{references}

\subsection{Appendix A: Effective trimer hopping Hamiltonian}

For the trimer the effective hopping Hamiltonian is
\begin{equation}
    H = \sum_{i,\sigma,\sigma'} K^0_{\sigma,\sigma'} \hat m_{i\sigma}^\dagger \hat m_{i\sigma'} + \sum_{i,\sigma,\sigma'} K^1_{\sigma,\sigma'} \hat m_{i+1,\sigma}^\dagger \hat m_{i\sigma'} + H.c.
\end{equation}
where the corresponding matrices for the five-state trimer propagation are
\begin{equation}
K^0 = 
\left(\begin{array}{ccccc}
0 & t_\mathrm{m} & 0 & 0 & 0 \\
t_\mathrm{m} & 0 & t_\mathrm{m} & t_0 & 0\\
0 & t_\mathrm{m} & 0 & 0 & 0 \\
0 & t_0 & 0 & 0& t_\mathrm{m} \\
0 & 0 & 0 & t_\mathrm{m} & 0
\end{array}\right)
\end{equation}
and
\begin{equation}
K^1 = 
\left(\begin{array}{ccccc}
0 & 0 & 0 & 0 & 0 \\
0 & 0 & 0 & 0 & 0\\
0 & 0 & 0 & t_\mathrm{m} & 0 \\
0 & 0 & 0 & 0& 0 \\
0 & 0 & 0 & 0 & 0
\end{array}\right),
\end{equation}
where the basis states correspond to configurations $\ket{\uparrow \downarrow \downarrow}$, $\ket{\updownarrow \downarrow}$, $\ket{\downarrow \uparrow \downarrow}$, $\ket{\downarrow \updownarrow}$, and $\ket{\downarrow \downarrow \uparrow}$.

For the three-state trimer, the matrices are
\begin{equation}
K^0 = 
\left(\begin{array}{ccc}
0 & t_\mathrm{m} & t_0\\
t_\mathrm{m} & 0 & 0\\
t_0 & 0 & 0\\
\end{array}\right)
\end{equation}
and
\begin{equation}
K^1 = 
\left(\begin{array}{ccc}
0 & 0 & 0\\
0 & 0 & t_\mathrm{m} \\
0 & 0 & 0\\
\end{array}\right),
\end{equation}
where the basis states correspond to configurations $\ket{\updownarrow\downarrow}$, $\ket{\downarrow \uparrow \downarrow}$, and $\ket{\downarrow \updownarrow}$.
The three-state trimer eigenstates are solutions of the characteristic equation
\begin{equation}
    \lambda^3 - \lambda\left[2t_\mathrm{m}^2 +t_0^2\right] + 2 t_\mathrm{m}^2 t_0 \cos k = 0.
\end{equation}

\end{document}